\documentclass[aps,reprint,superscriptaddress,floatfix]{revtex4-2}
\usepackage{amsmath,amssymb,graphicx,hyperref,booktabs,microtype,needspace}
\hypersetup{colorlinks=true,linkcolor=black,citecolor=black,urlcolor=blue,
 pdftitle={Stable Electroweak Strings and Nambu Monopoles Nested at Inert-Higgs Interfaces},pdfauthor={Muneto Nitta}}
\newcommand{\dd}{\mathrm d}\newcommand{\EM}{\mathrm{em}}\newcommand{\walloverlap}{\tfrac{\sqrt2}{12}}
\usepackage{xcolor}
\colorlet{BLUE}{blue}
\begin{document}
\title{Stable Electroweak Strings and Nambu Monopoles Nested at Inert-Higgs Interfaces}
\author{Muneto Nitta}
\affiliation{Department of Physics \& Research and Education Center for Natural Sciences, Keio University, 4-1-1 Hiyoshi, Yokohama, Kanagawa 223-8521, Japan}
\affiliation{International Institute for Sustainability with Knotted Chiral Meta Matter (WPI-SKCM$^2$), Hiroshima University, 1-3-1 Kagamiyama, Higashi-Hiroshima, Hiroshima 739-8531, Japan}
\date{29 October 2026}
\begin{abstract}
Electroweak $Z$ strings are generally unstable, and their Nambu-monopole endpoints lack topological protection in the bulk. Using degenerate inert and inert-like Higgs vacua as a solvable reference, we show that their interface supports both defects. The analytic interface carries a normalizable $SU(2)$ relative-orientation field. Its effective theory supports locally stable unit-flux $Z$ strings in a one-vacuum phase and fractional $Z$ strings in a two-vacuum phase. The fractional string carries an $S^2$ orientation field; a Nambu monopole is a second kink between symmetry-related, equal-tension $Z$ strings. We identify phases in which Nambu monopoles are confined or deconfined within the interface: in the confined phase, their $Z$ flux is concentrated into fractional strings, whereas in the deconfined phase, the monopole becomes a vortex within the interface and its $Z$ flux spreads radially through the interface. The electromagnetic magnetic field extends outside the interface in both phases. We determine analytic string profiles, effective-theory stability regions, and the unchanged electromagnetic magnetic charge across this confinement transition.
\end{abstract}
\maketitle

\noindent\textit{Introduction.} Could electroweak magnetic flux form persistent strings? Nambu's electroweak flux configurations and the $Z$ string have long provided concrete settings for this question \cite{nambu,vachaspati,review}. Such strings have been investigated as possible ingredients in electroweak baryogenesis \cite{baryogenesis} and the generation of primordial magnetic fields \cite{magnetic,helicity}. Their persistence is the difficulty. The bulk electroweak vacuum does not topologically protect a $Z$ string, and the familiar string is unstable at physical Standard Model couplings; its unstable modes have been mapped in detail \cite{james1992,james,goodband}. Its Nambu monopole endpoint likewise lacks protection from bulk vacuum topology. Proposed mechanisms have involved fields bound to the string \cite{vachaspatiwatkins}, strong external magnetic fields \cite{garriga}, thermal photons \cite{nagasawa}, and additional scalar condensates \cite{forgacs}. Fermionic effects have also been examined, with conclusions depending on the setup \cite{naculich,etohamada2024}.

These conclusions concern the minimal Higgs sector of the Standard Model, but neither gauge symmetry nor the observed Higgs boson requires the scalar sector to contain only one doublet. Two-Higgs-doublet models provide a simple, well-motivated extension, with richer vacuum structures and new possibilities for the electroweak phase transition \cite{2hdmreview,kanemuraewbg,aks,aksewbg,vacuumtopology}. Electroweak flux tubes and fractional strings \cite{earlyvortex,fluxtube,twohiggsvortex,etokurachi,na2hdm}, monopole junctions and string molecules \cite{topmonopole,monodynamics,monotopology,molecule}, and charged, CP-violating, or secondary walls \cite{membrane,cpsoliton,wallsim,wallconstraint,collapse,chargedwall,wallclassification} have been studied in these models. Superconducting strings, wall-bound condensates, and global monopoles add further possibilities \cite{superconducting2hdm,percolation,kinkyvortons,globalmonopoles,hopffibration}. In particular, binding two fractional strings can stabilize an electroweak flux tube in the bulk \cite{molecule}.

The inert doublet model, a simple, $\mathbb Z_2$-symmetric version of the two-Higgs-doublet model, provides a particularly useful setting for studying interfaces. Its discrete symmetry forbids Yukawa couplings of the second doublet; the standard inert vacuum admits a dark-matter candidate \cite{deshpande,barbieri,idmdm}, and its additional scalars are constrained by collider and precision-Higgs studies \cite{aokicollider,idmprecision,idmlimits}. They can also strengthen the electroweak phase transition \cite{idmewpt,gil,idmperturbativity}. Competing inert and inert-like minima permit multistep thermal histories \cite{ferreira,ginzburg,aokimultistep,aoki,fabian}, including studies of intermediate charge-breaking phases \cite{aokicharged}. Despite these extensive phenomenological studies, topological defects and their physical implications in the inert doublet model have received much less attention than in general two-Higgs-doublet models.

Here we show that the interface supports stable $Z$ strings within a controlled effective theory in a regime where the corresponding isolated bulk string is unstable. We use equal-depth vacua as an analytic reference and discuss the biased, moving interface relevant away from degeneracy. A normalizable relative-orientation field localized on the interface turns the string into a kink of the interface field; we derive its $Z$ flux and identify open regions of local stability within the interface effective theory. A second kink between symmetry-related fractional strings gives a topologically protected Nambu monopole. Across a transition between interface vacua, its $Z$ flux spreads through the interface while its outward electromagnetic magnetic charge remains unchanged.

\begin{figure*}[t]\centering
\includegraphics[width=\textwidth]{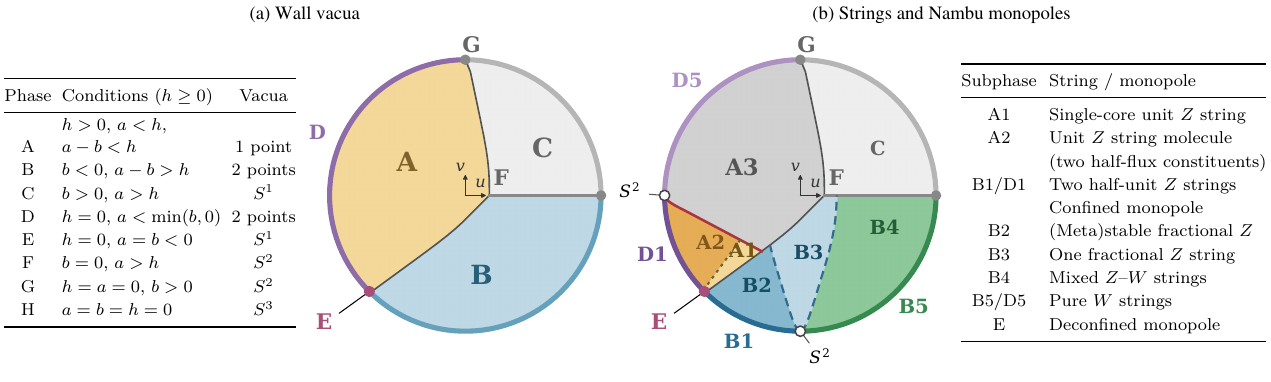}
\caption{Wall vacua and defects in coordinates $u=\rho\cos\theta$, $v=\rho\sin\theta$, with $-1\le u,v\le1$ and $u^2+v^2\le1$. Here $\rho=\log[1/(1-\alpha r)]/\log[1/(1-\alpha)]$, $r=\sqrt{a^2+b^2}/\sqrt{a^2+b^2+h^2}$, and $\theta=\arg(a+ib)$, with $\alpha=1-10^{-3}$. The rim has $h=0$ (exact inert $\mathbb Z_2$); H is the separate origin $a=b=h=0$. (a) Vacua of Eq.~(\ref{eq:wallpotential}) before electromagnetic identification: C is a charged circle, E a neutral circle. See Appendix~\ref{app:vacua}. (b) Strings at $g'=0$. Warm A1 has a single-core unit $Z$ string; A2 binds two half-flux constituents into a unit $Z$ string molecule. Gray A3 is charged-unstable, bounded by the red curve $q=\mu_0(\delta)$. B1/D1 have two degenerate half-unit-flux $Z$ strings; B2 stable/metastable fractional $Z$ strings; B3 one stable fractional $Z$ string; B4 mixed $Z$--$W$ strings; B5/D5 pure $W$ strings. Dashed B boundaries $a=\pm\pi h/2$ extrapolate from $|a|,h\ll-b$. Green B4/B5 and purple D5 support superconducting strings. Light-gray C/F/G lie outside these kink-string branches. Open circles indicate $S^2$ string-orientation moduli. The table includes locally stable strings and balanced/deconfined monopoles. In B1/D1, two fractional strings on opposite sides confine each monopole's $Z$ flux; at E it spreads radially within the wall (Fig.~\ref{fig:mono}).}\label{fig:phase}
\end{figure*}

\noindent\textit{Model.} Let $H_{1,2}$ be hypercharge-$1/2$ doublets, $r_i\equiv H_i^\dagger H_i$ and $\zeta\equiv H_1^\dagger H_2$. The bosonic part of the inert two-Higgs-doublet model Lagrangian is
\begin{align}
{\cal L}&=\sum_i|D_\mu H_i|^2-\frac14(W^a_{\mu\nu})^2-\frac14(Y_{\mu\nu})^2-V,\\
V={}&\sum_i\left(-m_i^2r_i+\frac{\lambda_i}{2}r_i^2\right)
+\lambda_3r_1r_2+\lambda_4|\zeta|^2\notag\\
&+\frac{\lambda_5}{2}(\zeta^2+\zeta^{*2})-m_{12}^2(\zeta+\zeta^*).\label{eq:model}
\end{align}
Here $D_\mu\equiv\partial_\mu-igW_\mu^a\sigma_a/2-ig'Y_\mu/2$, and all displayed scalar parameters are real. At $m_1^2=m_2^2=m^2$, $\lambda_1=\lambda_2=\lambda$, and $m_{12}^2=0$, the interchange $H_1\leftrightarrow H_2$ makes $(v\xi_0,0)$ and $(0,v\xi_0)$ degenerate, where $v^2=m^2/\lambda$ and $\xi_0=(0,1)^T$. We work where these inert and inert-like vacua are minima. The bulk symmetry assignments follow Refs.~\cite{symmetry2012,symmetry2020}. A sufficiently small exchange-preserving $m_{12}^2$ maintains their degeneracy but induces a secondary condensate; ``inert'' then refers to their continuation. We use degenerate vacua as an analytic reference; a small bias lifting the inert-like minimum is discussed below.

\noindent\textit{Domain wall and its effective theory.} At the custodial reference point $g'=\lambda_4=\lambda_5=m_{12}^2=0$, choose units $m^2=\lambda=1$ and $\lambda_3=3$. Custodial symmetry is shared with the enlarged scalar-symmetry frameworks of Ref.~\cite{maxsym}, but this reference potential need not have their maximal $SO(5)$ symmetry. With $z$ the spatial coordinate normal to the interface and $(x,y)$ the two coordinates within it, an exact wall is
\begin{equation}
\begin{gathered}
H_1=f_-(z)\xi_0,\qquad H_2=f_+(z)U\xi_0,\\
f_\pm(z)=\tfrac12[1\pm\tanh(z/\sqrt2)],\quad U\in SU(2).
\end{gathered}\label{eq:wall}
\end{equation}
Its gauge field vanishes and $T_0=2\sqrt2/3$. A common rotation is gauge; the relative orientation $U$ has finite kinetic norm after gauge relaxation. The reference wall has translation and three orientation zero modes. For the remaining massive physical fluctuations we prove $M_\perp^2\ge\min(g^2/4,1)>0$, excluding the retained bulk photon (\hyperref[app:metric]{Appendix A}). The relative field is the electroweak counterpart of non-Abelian wall clouds \cite{earlywall,wallskyrmion,cloud,jwall,josephson,host}.

In the moduli approximation \cite{manton,solitoneft}, we promote $U$ to a slowly varying wall field and integrate over $z$. The leading effective theory is an $SU(2)$ principal chiral model, with $SU(2)\simeq S^3$, perturbed by a potential. The unbroken electromagnetic group acts by $U\mapsto e^{i\omega\sigma_3/2}Ue^{-i\omega\sigma_3/2}$; the photon thus enters through $\mathcal D_\alpha U\equiv\partial_\alpha U-ieA_\alpha[\sigma_3/2,U]$, with $e\equiv gg'/\sqrt{g^2+g'^2}$. Its wall Lagrangian is
\begin{align}
{\cal L}_w&=\frac{K_g}{2}\operatorname{tr}[({\mathcal D}_\alpha U)^\dagger {\mathcal D}^\alpha U]-V_w(U),\label{eq:LU}\\
V_w(U)&=\tfrac{\sqrt2\lambda_4}{48}[2+\operatorname{tr}(\sigma_3U\sigma_3U^\dagger)]\notag\\
&+\tfrac{\sqrt2\lambda_5}{48}[(\operatorname{tr}U)^2+(\operatorname{tr}(\sigma_3U))^2]-\tfrac{1}{\sqrt2} m_{12}^2\operatorname{tr}U.\label{eq:VU}
\end{align}
Here $\alpha=t,x,y$, and $\sigma_3$ specifies the reference vacuum direction, $\xi_0\xi_0^\dagger=(\mathbf1-\sigma_3)/2$. The first-order potential uses the reference-wall overlap integrals $\int\dd z\,f_-^2f_+^2=\sqrt2/12$ and $\int\dd z\,f_-f_+=1/\sqrt2$, whose values are inserted explicitly throughout.
A slowly varying $U$ induces $W_\alpha=-iq_g(z)(\partial_\alpha U)U^\dagger/g$, $W_z=0$. The dimensionless profile $q_g(z)$ depends on the gauge coupling $g$. The profile minimizes the gauge-gradient functional below with $q_g(-\infty)=0$, $q_g(+\infty)=1$. Evaluating it on this profile gives the positive kinetic coefficient
\begin{equation}
K_g=\int\dd z\,[f_-^2q_g^2+f_+^2(1-q_g)^2+2(q_g')^2/g^2].\label{eq:K}
\end{equation}
where $q_g'\equiv\dd q_g/\dd z$ (\hyperref[app:metric]{Appendix A}). 
For the vacuum and soliton analysis, we parameterize
\begin{equation}
U=\begin{pmatrix}u_0-iu_3&c\\-c^*&u_0+iu_3\end{pmatrix},\quad
u_0^2+u_3^2+|c|^2=1.\label{eq:U}
\end{equation}
The same principal chiral model then reads
\begin{align}
{\cal L}_{w}&=K_g[(\partial u_0)^2+(\partial u_3)^2+|{\mathcal D}c|^2]-V_w,\label{eq:Lw}\\
V_w&=\walloverlap (au_0^2+bu_3^2-2hu_0),\label{eq:wallpotential}\\
a&\equiv\lambda_4+\lambda_5,\quad b\equiv\lambda_4-\lambda_5,\quad h\equiv6m_{12}^2.
\end{align}
The derivatives in Eq.~(\ref{eq:Lw}) are along the wall. Electromagnetism acts only on $c$: $c\mapsto e^{i\omega}c$, ${\mathcal D}_\alpha c\equiv(\partial_\alpha-ieA_\alpha)c$, while $u_0,u_3$ are neutral. The three-dimensional photon retains its bulk Maxwell term. For neutral fields $c=0$, gauge dressing at nonzero $g'$ additionally replaces $K_g$ by $K_Z\equiv K_{g_Z}<K_g$, with $g_Z\equiv\sqrt{g^2+g'^2}$.

The wall vacua A--H are classified in Fig.~\ref{fig:phase}(a). The string solutions below concern A and B, and the monopole transition passes through B--E--D.

\noindent\textit{$Z$ strings as kinks inside the wall.} We now show that a kink of the wall field carries $Z$ flux and identify its stable branches. Set $u_0+iu_3=e^{i\varphi(x)}$, $c=0$, for a straight string along $y$. Its neutral energy per unit length is
\begin{equation}
\begin{aligned}
E_n=\int\dd x\big[&K_Z(\varphi')^2+\walloverlap (a\cos^2\varphi\\
&+b\sin^2\varphi-2h\cos\varphi)-V_{\min}\big],
\end{aligned}\label{eq:neutral}
\end{equation}
where $\varphi'\equiv\dd\varphi/\dd x$ and $V_{\min}$ is the vacuum value of $V_w$. The relaxed neutral gauge field is $Z_x=-2q_Z(z)\varphi'(x)/g_Z$, with $q_Z(-\infty)=0$ and $q_Z(+\infty)=1$. Therefore the flux through the transverse $(z,x)$ plane is
\begin{equation}
\Phi_Z\equiv\int\dd z\,\dd x\,F^Z_{zx}
=-\frac{2\Delta\varphi}{g_Z}.\label{eq:flux}
\end{equation}
Here $Z_\mu\equiv(gW_\mu^3-g'Y_\mu)/g_Z$ in the neutral gauge, and $F^Z_{zx}\equiv\partial_zZ_x-\partial_xZ_z$. Equation~(\ref{eq:flux}) identifies the wall kink as a $Z$ string; a full $2\pi$ kink has the unit flux magnitude $4\pi/g_Z$ (\hyperref[app:flux]{Appendix B}).

\noindent\textbf{A phase: unit-flux strings.} Put $\delta\equiv(b-a)/h>-1$, $q\equiv b/h<1+\delta$, and $X\equiv\sqrt{\walloverlap h/K_Z}(x-x_0)$. The exact neutral unit-flux profile is a $2\pi$ double sine-Gordon (DSG) kink \cite{rossshort,amarishort},
\begin{equation}
\varphi(X)=\pi+2\arctan\frac{\sinh(\sqrt{1+\delta}\,X)}{\sqrt{1+\delta}}.\label{eq:Aprofile}
\end{equation}
At $x=x_0$, $\varphi=\pi$, and the profile obeys $\varphi(x)+\varphi(2x_0-x)=2\pi$. Equation~(\ref{eq:flux}) therefore assigns each half of this symmetric solution a phase change $\Delta\varphi=\pi$ and half-unit flux $|\Phi_Z|=2\pi/g_Z$. For $\delta>1$, $\varphi=\pi$ is a local minimum along the neutral path, lying $\sqrt2 h/3$ above the true vacuum. The two half-flux cores thus form a confined molecule; at $h>0$ neither is an isolated finite-energy kink.
Unlike the familiar unstable bulk string, this kink is locally stable in an open parameter region. Expanding in the charged field $c$ at $g'=0$ gives $\mathcal O_c\psi=\omega^2\psi$ for either real component at zero longitudinal momentum, with
\begin{equation}
\mathcal O_c=\frac{\sqrt2 h}{12K_g}\big[-\partial_X^2-(\varphi_\delta')^2+\delta\cos^2\varphi_\delta+\cos\varphi_\delta-q\big].
\label{eq:chargedA}
\end{equation}
Here $\varphi_\delta$ is Eq.~(\ref{eq:Aprofile}), and the prime means $\dd/\dd X$. If $\mu_0(\delta)$ is the lowest bracket eigenvalue at $q=0$, the charged gap is $(\walloverlap h/K_g)[\mu_0(\delta)-q]$. The neutral operator is nonnegative, so $q<\mu_0(\delta)$ tests all three wall-field directions (\hyperref[app:metric]{Appendix A}). For example, $\delta=0$, $q=-4$ gives the positive lowest eigenvalue $\sqrt2h/(12K_g)$ analytically. Figure~\ref{fig:phase}(b) separates stable single-core strings A1 ($\delta<1$), molecules of two fractional $Z$ strings A2 ($\delta>1$), and A3, where a charged instability opens an unwinding channel toward the homogeneous vacuum. At fixed $\delta,q$ with a strict stability margin, $h\to0^+$ separates the kink scale from $M_\perp$. Weak nonzero $g'$ requires explicit photon dynamics and corrections smaller than this margin; we do not determine the full finite-$g'$ boundary. This mechanism differs from bulk string-molecule stabilization \cite{molecule}.

Along a path through A1 and A2 at fixed $a<b<0$, decreasing $h$ increases $\delta$: the single-core A1 string becomes an A2 molecule and, as $h\to0^+$, dissociates into two independent half-unit strings in D1, discussed below. The intervening vacuum becomes degenerate, its confining energy density $\sqrt2h/3$ vanishes, and the separation diverges. This constituent deconfinement differs from monopole deconfinement at E and charged unwinding in A3.

\noindent\textbf{B phase: fractional strings.} In contrast to the A phase, the B phase admits two types of fractional $Z$-string solution joining its two wall vacua. Each is a topological kink supported by the homotopy group $\pi_0$, since its asymptotic wall vacua are disconnected. Their stable orientations are determined by a second effective theory. At $a=h=g'=0$, $b<0$, the fractional kink has an exact $S^2$ family:
\begin{equation}
\begin{gathered}
u_3=\cos\theta,\quad(\Re c,\Im c,u_0)=\boldsymbol n\sin\theta,\quad\boldsymbol n^{\,2}=1,\\
\theta=2\arctan e^{m_B(x-x_0)},\quad m_B\equiv\sqrt{\walloverlap (-b)/K_g}.
\end{gathered}\label{eq:S2}
\end{equation}
Here $m_B$ is the inverse fractional-string width in the $x$ direction within the interface, perpendicular to the string along $y$, and $\boldsymbol n=(n_1,n_2,n_3)\in S^2$. Promoting $\boldsymbol n$ to a field on the string yields an $O(3)$ nonlinear sigma model on $S^2$. With $\beta=t,y$, $G=2K_g/m_B$ and ${\mathcal D}_\beta\boldsymbol n\equiv\partial_\beta\boldsymbol n-eA_\beta(\hat{\boldsymbol e}_3\times\boldsymbol n)$, small $a,h,g'^2$ give the effective theory
\begin{equation}
\begin{aligned}
{\cal L}_{\rm str}&=G({\mathcal D}_\beta\boldsymbol n)\cdot({\mathcal D}^\beta\boldsymbol n)-V_{\rm str}(n_3),\\
V_{\rm str}(n_3)&=\mathrm{const}-Dn_3^2-Hn_3,\quad -1\le n_3\le1,\\
D&=C-\frac{\sqrt2 a}{6m_B},\quad H=\frac{\pi\sqrt2 h}{6m_B}.
\end{aligned}\label{eq:Bpot}
\end{equation}
Here $C$ is the hypercharge-induced anisotropy coefficient, given to leading order for weak $g'\ne0$ by
\begin{equation}
C=\frac{2g'^2}{g^4}\sqrt{\frac{\sqrt2(-b)}{3K_g}}\int\dd z\,(q_g')^2+O(g'^4)>0.\label{eq:Caxis}
\end{equation}
Here $C$ favors $Z$ orientations and vanishes at $g'=0$ (\hyperref[app:orientation]{Appendix B}). The net anisotropy $D$ is easy-axis for $D>0$, easy-plane for $D<0$. Equation~(\ref{eq:Bpot}) classifies the tensions: B1 ($H=0,D>0$) has degenerate $Z$ strings; B2 ($0<H<2D$) stable/metastable $Z$ strings; B3 ($H>0$, $-H/2<D<H/2$) one stable $Z$ string; B4 ($H>0$, $D<-H/2$) a circle of mixed $Z$--$W$ strings; B5 ($H=0,D<0$) pure $W$ strings. Mixed $Z$--$W$ and pure $W$ strings are superconducting: their charged $U(1)$ phase couples to the bulk photon and permits longitudinal currents. These interface-field kinks differ from the Higgs-family strings of Ref.~\cite{superconducting2hdm}.

The neutral branches in B1 and B2 can now be compared explicitly. With $d\equiv a-b>h$ and $\varphi_0\equiv\arccos(h/d)$, they follow the short path $-\varphi_0\to\varphi_0$ and the complementary path $\varphi_0\to2\pi-\varphi_0$. These are the short and long double sine-Gordon walls of Refs.~\cite{rossshort,amarishort}; here they carry electroweak flux inside the interface. With $r\equiv h/d$ and $\mathcal T\equiv\sqrt{K_Z\walloverlap d}$, their tensions and $Z$-flux magnitudes are
\begin{align}
T_{\rm short}&=4\mathcal T(\sin\varphi_0-r\varphi_0),\quad
|\Phi_{Z,\rm short}|=\tfrac{4\varphi_0}{g_Z},\label{eq:Btensions}\\
T_{\rm long}&=4\mathcal T[\sin\varphi_0+r(\pi-\varphi_0)],\quad
|\Phi_{Z,\rm long}|=\tfrac{4(\pi-\varphi_0)}{g_Z}.\notag
\end{align}
Their flux magnitudes sum to one unit, $4\pi/g_Z$, 
so that $\Delta T\equiv T_{\rm long}-T_{\rm short}=4\pi r\mathcal T$. On the B1 line ($h=0$, $D>0$), the two neutral strings have equal tension and half-unit flux $|\Phi_Z|=2\pi/g_Z$:
\begin{equation}
T_{\rm frac}=2\sqrt{\frac{\sqrt2 K_Z(a-b)}{3}},\quad
\ell_{\rm frac}=\frac{4K_Z}{T_{\rm frac}}.
\label{eq:frac}\end{equation}
Here $\ell_{\rm frac}$ is the kink width within the wall, the inverse mass scale in its profile. \hyperref[app:metric]{Appendix A} verifies charged stability for $a<0$ at $h=g'=0$. Taking $a=-\alpha\varepsilon$, $b=-\beta\varepsilon$ with fixed $0<\alpha<\beta$ and sufficiently small $\varepsilon$ separates the string scale and its positive charged gap from $M_\perp$. At finite $h$ and $g'=0$, $a<-2h-h^2/d$ is a sufficient condition for both neutral branches to remain locally stable (\hyperref[app:metric]{Appendix A}). 

At $h=0$, the neutral potential above its minimum is $\walloverlap(a-b)\cos^2\varphi$. Concatenating the two increasing B1 kinks gives unit flux, but the sine-Gordon kinks repel without a confining vacuum-energy cost. Thus this unit-flux configuration dissociates into two half-unit strings. B1 has $b<a$; reaching it from A crosses $h=a-b$, where the wall vacuum bifurcates, rather than following the A2 limit.

\noindent\textbf{D phase.} The D phase has the same orientation construction after exchanging $u_0$ and $u_3$, or equivalently $a$ and $b$ at $h=0$. Its easy-axis branch, which we call D1, has two balanced half-unit $Z$ strings; its easy-plane branch D5 has pure $W$ strings (\hyperref[app:orientation]{Appendix B}). Thus D1 is the counterpart of B1, and Eq.~(\ref{eq:frac}) holds with $a$ and $b$ exchanged.

\begin{figure*}[t]\centering\includegraphics[width=.88\textwidth]{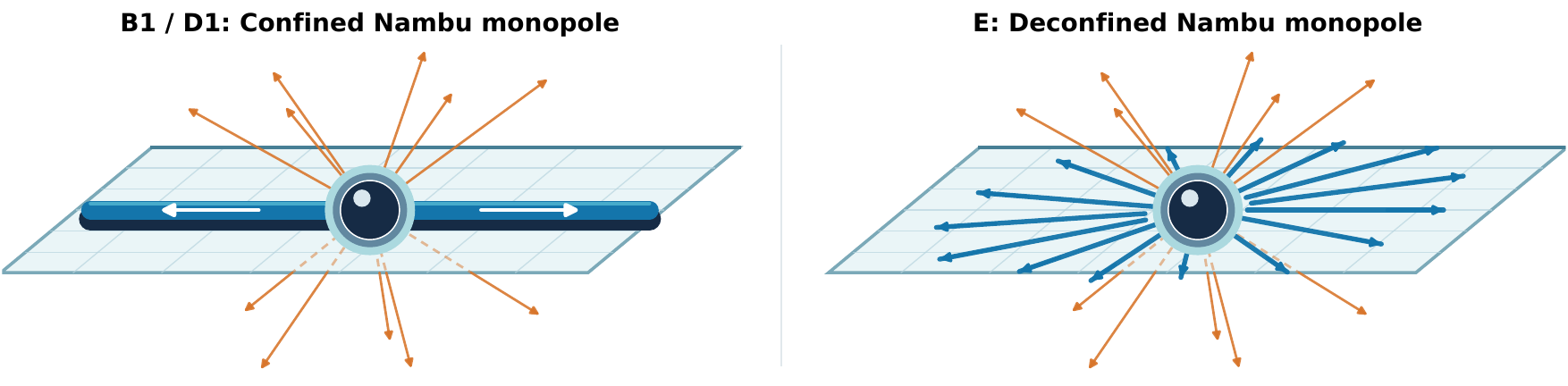}
\caption{Interface Nambu monopoles. In B1/D1 (left), two half-unit strings confine the outward $Z$ flux to two branches, each of magnitude $2\pi/g_Z$. In E (right), the vortex spreads $Z$ flux radially within the wall, with total outward magnitude $4\pi/g_Z$. Blue arrows show $Z$ flux in the interface; orange arrows show outward electromagnetic flux in the bulk. This is a schematic; the flux matching is analytic (\hyperref[app:flux]{Appendix B}).}\label{fig:mono}\end{figure*}

\Needspace{3\baselineskip}\noindent\textit{Confined and deconfined Nambu monopoles.} Conventional bulk Nambu monopoles lack topological protection and are attached to generally unstable electroweak $Z$ strings. Here the wall supports topologically protected monopoles that are stable within the effective theory: nested kinks in B1/D1 and a $\pi_1(S^1)$ vortex without attached $Z$ strings in the E phase. 

Equal string tensions follow from symmetry. At $h=0$, the CP-conserving model has the exact inert $\mathbb Z_2$, $H_2\mapsto-H_2$. In B1, its product with CP acts as $\zeta\mapsto-\zeta^*$; in D1, CP acts as $\zeta\mapsto\zeta^*$. These transformations preserve the respective asymptotic wall vacua and exchange the two string branches, enforcing equal tensions, as in the symmetry-related bulk junctions of Refs.~\cite{topmonopole,monodynamics}. With exact inert $\mathbb Z_2$ symmetry ($m_{12}^2=0$), the balanced monopoles in B1/D1 are stable over finite ranges of the remaining couplings within the effective theory. These regions appear on the rim of Fig.~\ref{fig:phase} because its interior represents soft $\mathbb Z_2$ breaking, $m_{12}^2\ne0$.

The two stable $Z$ orientations in B1 or D1 permit a second kink, now along the string. Set $n_3=\cos\alpha$, $n_1+in_2=\sin\alpha e^{i\chi}$, and ${\cal D}_\beta\chi\equiv\partial_\beta\chi-eA_\beta$ for $\beta=t,y$. In B1, Eq.~(\ref{eq:Bpot}) gives
\begin{equation}
{\cal L}_{1+1}=G[(\partial\alpha)^2+\sin^2\alpha({\cal D}\chi)^2]-D\sin^2\alpha.
\label{eq:orient}
\end{equation}
For a static configuration with constant $\chi$ and vanishing longitudinal photon, its pole-to-pole kink is
\begin{equation}
\begin{aligned}
\alpha(y)&=2\arctan e^{m_M(y-y_0)},\quad m_M\equiv\sqrt{D/G},\\
M_M&=4Gm_M=4\sqrt{GD}.
\end{aligned}\label{eq:monokink}
\end{equation}
This kink is a Nambu monopole joining the two symmetry-related half-unit $Z$ strings and emitting unit outward $Z$ flux [Fig.~\ref{fig:mono} (left)]. Here $m_M$ is the inverse width along the string, and $M_M$ the kink energy above the string background. This effective theory requires $m_M\ll m_B\ll M_\perp$, or $-a/(a-b)\ll1$ at $h=g'=0$, with corrections below the stability margins. The massless photon propagates in the bulk and is retained as a dynamical field coupled to the defects. At $g'=0$, $\chi$ is a normalizable $S^1$ phase modulus. 

At E, $h=\lambda_5=0$ restores relative-phase $U(1)$ symmetry, spontaneously broken on the wall. Its neutral vacuum circle supports a $\pi_1(S^1)$ vortex with a charged core, $u_0+iu_3\sim e^{i\vartheta}$, where $\vartheta=\arg(x+iy)$. This vortex spreads $Z$ flux radially within the wall (Fig.~\ref{fig:mono}, right), with far-wall energy $2\pi K_Z\log(L/\xi)$, where $L$ and $\xi$ are infrared and core scales. The monopole is deconfined within the wall but remains bound to it. As E is approached, the fractional-string tension vanishes and its width $\ell_{\rm frac}$ diverges. The effective theory on the string is then no longer valid. The monopole must instead be described as a vortex in the wall effective theory.

The electromagnetic charge survives this transition. A regular hypercharge connection with no fundamental hypermagnetic source satisfies $F^Y=\cos\theta_WF^{\EM}-\sin\theta_WF^Z$. On a closed surface surrounding the junction or vortex, $\int F^Y=0$ and the outward $Z$ flux is $4\pi/g_Z$. Consequently (\hyperref[app:flux]{Appendix B}),
\begin{equation}
\Phi_{\EM}=t_W\Phi_Z^{\rm out}=\frac{4\pi s_W}{g},\quad
\mathbf B_{\EM}\sim\frac{s_W\mathbf r}{gr^3}.\label{eq:charge}
\end{equation}
where $s_W\equiv\sin\theta_W=g'/g_Z$, $t_W\equiv\tan\theta_W$, $\Phi_{\EM}$ is the outward electromagnetic magnetic flux, and $\mathbf r=(x,y,z)$ is measured from the monopole. The sign reverses for an antimonopole.

Unstable monopole configurations occur in other phases. In A1/A2, a unit $Z$ string can terminate where its winding unwinds through charged field directions. The endpoint carries Nambu-monopole magnetic charge and is pulled by string tension; its full field profile remains to be calculated. In B2, the two string tensions differ, so their junction accelerates under the resulting force \cite{monodynamics}. At $g'=0$, Eq.~(\ref{eq:Bpot}) admits unstable monopole--antimonopole pairs in B2/B3. The field $\boldsymbol n=(\sin\alpha,0,\cos\alpha)$ follows a great circle from the north pole through the south pole and back, with $\alpha:0\to2\pi$. Its profile is the DSG kink~\cite{rossshort,amarishort} in Eq.~(\ref{eq:Aprofile}), with $\delta=2D/H$ and $X=\sqrt{H/(2G)}(y-y_0)$, for $H>0$, $D>-H/2$. A rotation of the great circle gives the zero mode $\delta n_2\propto\sin\alpha$, with one internal node. Since the ground state of the one-dimensional fluctuation operator has no nodes, it lies below this zero mode and has a negative eigenvalue. The pair is therefore a sphaleron that can unwind out of the great circle and annihilate.

\noindent\textit{Illustrative scales.} Taking $m=88.46\,\mathrm{GeV}$, $\lambda=0.2581$, $\lambda_3=3\lambda$, $g=0.653$, and $g'=0.358$, the A1 point $\lambda_4=-0.04\lambda$, $\lambda_5=0$, $m_{12}^2=13.04\,\mathrm{GeV}^2$ gives $T_{\rm unit}\simeq1.1\times10^4\,\mathrm{GeV}^2$. At the distinct B1 point $(\lambda_4,\lambda_5)=(-0.051,0.049)\lambda$, $m_{12}^2=0$, Eqs.~(\ref{eq:frac}) and (\ref{eq:monokink}) give $T_{\rm frac}\simeq8.7\times10^3\,\mathrm{GeV}^2$ and $M_M\simeq0.52\,\mathrm{TeV}$.

\noindent\textit{Expanding interfaces.} The Yukawa sector distinguishes the doublets, so their degeneracy is not protected. A bias $\Delta V\equiv V_{\rm IL}-V_{\rm I}>0$ can retain a metastable inert-like minimum. At $m_{12}^2=0$, define $\kappa\equiv\lambda_3+\min(0,a,b)$ and $\rho\equiv(m_1^2/\sqrt{\lambda_1})/(m_2^2/\sqrt{\lambda_2})$, with $m_i^2>0$. The tree-level condition $1<\rho<\kappa/\sqrt{\lambda_1\lambda_2}$ makes the inert vacuum deeper while keeping all physical scalar curvatures positive at both vacua \cite{ferreira,ginzburg}; it continues to small mixing (Appendix~\ref{sm:app:bias}). For small bias, the $U$ theory continues on a locally comoving bubble patch if curvature $R^{-1}$ and acceleration $\sim\Delta V/T_{\rm wall}$ are below $M_\perp$, profile shifts are perturbative, and evolution is adiabatic on defect relaxation scales. Its defects persist quasistatically while their stability margins remain positive. A sweeping bubble may capture a bulk string or string--monopole complex before it unwinds; capture competes with unwinding and radiation and requires time-dependent dynamics.

\begin{acknowledgments}
\noindent\textit{Acknowledgments.} This work was supported in part by JSPS KAKENHI Grants No. JP22H01221 and No. JP23K22492, and by WPI-SKCM$^2$ at Hiroshima University.
\end{acknowledgments}

\appendix
\section{String stability in the effective theory}\label{app:metric}
We establish two conditions for the string-stability claims: a gap for the omitted wall fluctuations and stability within the retained wall theory. We proceed from gauge relaxation to the reference-wall gap, string fluctuations, and control of the reduction. Unit conversions are collected at the end.

\noindent\textit{Wall kinetic coefficient.} Gauge relaxation fixes the finite kinetic norm of the relative orientation. For a slowly varying relative rotation, define $\Omega_\alpha\equiv-i(\partial_\alpha U)U^\dagger$ and take $W_\alpha=q_g\Omega_\alpha/g$, $W_z=0$, where $W_\alpha=W_\alpha^a\sigma_a/2$. Its scalar gradients are proportional to $q_gf_-$ and $(1-q_g)f_+$; the gauge field strength adds the last term of Eq.~(\ref{eq:K}). Minimization gives
\begin{equation}
-\frac{2}{g^2}q_g''+(f_-^2+f_+^2)q_g=f_+^2.
\end{equation}
Here $q_g'\equiv\dd q_g/\dd z$. The conditions $q_g(-\infty)=0$ and $q_g(+\infty)=1$ cancel the covariant orientation gradients in the left and right bulk vacua, respectively. A smooth interpolating trial profile has finite norm, and the positive minimum defines $K_g$. This is the induced gauge-profile equation, distinct from the scalar wall equations.

\noindent\textit{Gap of the reference wall.} We first bound the omitted wall modes, starting with Higgs-amplitude fluctuations. For the reference-wall stability test use $m^2=\lambda=1$, $\lambda_3=3$, $g>0$, and $g'=a=b=h=0$. Write $H_i=(f_i+\eta_i)\xi_0$, with $f_{1,2}=\tfrac12[1\mp\tanh(z/\sqrt2)]$ and real $\eta_i$. The real-amplitude Hessian is
\begin{equation}
\begin{aligned}
\mathcal H_{\rm amp}&=-\partial_z^2\mathbf1+
\begin{pmatrix}v&6f_1f_2\\6f_1f_2&v\end{pmatrix},\\
v&=-1+3(f_1^2+f_2^2).
\end{aligned}\label{eq:ampHessian}
\end{equation}
The combinations $\eta_\pm\equiv(\eta_1\pm\eta_2)/\sqrt2$ diagonalize it into $L_+=-\partial_z^2+2$ and $L_-=A^\dagger A$, with $A=\partial_z+\sqrt2\tanh(z/\sqrt2)$. The translation is its only zero mode. Its partner satisfies $AA^\dagger=B_0^\dagger B_0+3/2$, with $B_0=\partial_z+\tanh(z/\sqrt2)/\sqrt2$, so its nonzero spectrum is bounded below by $3/2$.

Next consider angular Higgs fluctuations coupled to the normal gauge field; the $\eta_i$ now denote angular fluctuations. For each of the three real angular Higgs--normal-gauge sectors, use $\Psi=(\eta_1,\eta_2,\nu)^T$ with kinetic norm $\int\dd z\,|\Psi|^2$, put $m_W=g/\sqrt2$ and $\varrho_i=f_i'/f_i$, and work at zero wall momentum. The normal quadratic energy is $Q_0=\int\dd z\sum_i|\eta_i^{\prime}-\varrho_i\eta_i+m_Wf_i\nu|^2$. Orthogonality to gauge orbits imposes $\nu'+m_W(f_1\eta_1+f_2\eta_2)=0$. Adding the square of this vanishing condition gives the gauge-fixed operator $\mathcal H_{\rm gf}=D^\dagger D$, where
\begin{equation}
D=\begin{pmatrix}
\partial_z-\varrho_1&0&m_Wf_1\\
0&\partial_z-\varrho_2&m_Wf_2\\
m_Wf_1&m_Wf_2&\partial_z
\end{pmatrix}.
\label{eq:end-gap-factor}
\end{equation}
A mass gap requires a positive bound on the partner operator. Let $t=m_W^2$, $F=(f_1,f_2)^T$, $N=(f_2,f_1)^T$, $S=f_1^2+f_2^2$, and $A_0=\operatorname{diag}(2f_2,2f_1)$. The exact profiles give
\begin{equation}
\begin{aligned}
DD^\dagger&=\operatorname{diag}(-\partial_z^2\mathbf1_2+V_P,L_V),\\
V_P&=A_0+tFF^T,\qquad L_V=-\partial_z^2+tS.
\end{aligned}
\label{eq:end-gap-partner}
\end{equation}
Since $f_i\ge0$ and $f_1+f_2=1$, $A_0+FF^T=\mathbf1_2+NN^T$. For $0<t\le1$, $V_P-t\mathbf1_2=tNN^T+(1-t)A_0\ge0$; for $t\ge1$, $V_P-\mathbf1_2=NN^T+(t-1)FF^T\ge0$. Also $S\ge1/2$. Hence
\begin{equation}
DD^\dagger\ge c_g\mathbf1_3,\qquad c_g=\min(g^2/4,1)>0.
\label{eq:end-gap-bound}
\end{equation}
The nonzero spectra of $D^\dagger D$ and $DD^\dagger$ coincide. We now identify physical zero modes and remove gauge redundancy. To apply this bound, let $\mathcal H_0$ be the unfixed angular Hessian and $R\alpha=(m_Wf_1\alpha,m_Wf_2\alpha,-\alpha')^T$. The horizontal condition is $R^\dagger\Psi=0$; gauge invariance gives $\mathcal H_0R=0$, so this subspace is invariant under $\mathcal H_{\rm gf}=\mathcal H_0+RR^\dagger$. Also $R^\dagger R=L_V>0$, excluding a normalizable residual gauge zero mode. The zero-mode equations imply $(\eta_1/f_1-\eta_2/f_2)'=0$; the horizontal condition then uniquely fixes the gauge response. There is one relative-rotation zero mode per real sector. Removing these modes and the radial translation leaves $M_\perp^2\ge c_g$: the radial gap is $3/2$ and tangential gauge modes obey $L_V\ge g^2/4$. Custodial symmetry makes the bound uniform in constant $U$. The massless hypercharge field is retained separately. This is a gap of the flat reference wall, not the Hessian of a wall containing a defect.

\noindent\textit{Stability of the wall strings.} We test neutral and charged string fluctuations in the retained wall theory. For a string in the A phase, the monotone profile in Eq.~(\ref{eq:Aprofile}) has a nodeless translation mode. Its neutral Hessian factors as
\begin{equation}
\mathcal O_n=(-\partial_x-w)(\partial_x-w),\qquad
w=\varphi''/\varphi',
\end{equation}
so it has no negative eigenvalue. The constrained charged expansion $u_0+iu_3=\sqrt{1-|c|^2}e^{i\varphi}$ gives Eq.~(\ref{eq:chargedA}) for either real component of $c$, including the metric term $-K_g(\varphi')^2|c|^2$. Hence $q<\mu_0(\delta)$ tests all three wall-field directions. At $\delta=0$, the dimensionless $q=0$ operator reduces to $-\partial_X^2+1-6\operatorname{sech}^2X$, giving the analytic check $\mu_0(0)=-3$. Momentum along the string adds $k_y^2$ to each eigenvalue.

For a string in the B phase at $h=g'=0$, put $m^2=\sqrt2(a-b)/(12K_g)$ and $A_B=\partial_x+m\tanh[m(x-x_0)]$. The charged operator is
\begin{equation}
\mathcal O_{B,c}=A_B^\dagger A_B-\frac{\sqrt2 a}{12K_g}.
\label{eq:end-Bcharged}
\end{equation}
The neutral short-branch profile obeys $\sin\varphi=\tanh[m(x-x_0)]$, $\cos\varphi=\operatorname{sech}[m(x-x_0)]$. Since $A_B$ annihilates the latter function, the lowest charged eigenfunction is $\operatorname{sech}[m(x-x_0)]$, with eigenvalue $-\sqrt2 a/(12K_g)$. The neutral operator is $A_B^\dagger A_B$. Thus $a<0$ gives local stability; $a=0$ restores two orientation zero modes. For finite $h$, $\mathcal O_c-\mathcal O_n=\sqrt2[-a-h^2/d+2h\cos\varphi]/(12K_g)$, so $a<-2h-h^2/d$ guarantees positivity for both branches. The D phase exchanges $a,b$.

\noindent\textit{Control of the reduction.} We finally bound corrections from mixing with omitted modes. These light-sector results are controlled when the defect widths exceed the massive transverse lengths. For $a=-\alpha\varepsilon$, $b=-\beta\varepsilon$, $0<\alpha<\beta$ fixed, both $m^2=\sqrt2(\beta-\alpha)\varepsilon/(12K_g)$ and the charged gap $m_c^2=\sqrt2\alpha\varepsilon/(12K_g)$ are $O(\varepsilon)$. Eliminating an orthogonal heavy block gives
\begin{equation}
\mathcal H_{\rm eff}=\mathcal H_{\ell\ell}-B\mathcal H_{hh}^{-1}B^\dagger,
\qquad \mathcal H_{hh}\gtrsim M_\perp^2.
\end{equation}
Here $\mathcal H_{hh}$ contains only massive physical transverse modes; symmetry zero modes and the photon are retained separately. The order-one-derivative gauge response has already been minimized into $K_g$. The remaining normal residual and its linearization are $O(p^2,\varepsilon)$, giving $B=O(\varepsilon)$ for light momenta $p^2=O(\varepsilon)$. Thus the correction is $O(\varepsilon^2/M_\perp^2)$, below strict $O(\varepsilon)$ light gaps at fixed ratios. This is a derivative-expansion argument, with gauge fixing and projection as above, rather than a full defect Hessian calculation. At finite $g'$ the photon remains explicit, and gauge corrections must remain below the stability margins. Replacing $K_g$ by $K_Z$ alone does not determine the charged boundary or establish stability at physical electroweak couplings.

\noindent\textit{Restoring physical units.} These conversions give the illustrative scales quoted in the main text. To restore dimensions, put $w=mz$, $\gamma=g/\sqrt\lambda$, and $\widehat h=6m_{12}^2/m^2$. Then $K_g=(m/\lambda)K(\gamma)$, where $K(\gamma)$ is Eq.~(\ref{eq:K}) in reference units; $K_Z$ uses $\gamma_Z=g_Z/\sqrt\lambda$. The illustrative values are $K_g=169.915\,\mathrm{GeV}$ and $K_Z=152.362\,\mathrm{GeV}$. With $P=\sqrt2m^3/(12\lambda)$, the $\delta=0$ first integral gives $T_{\rm unit}=16\sqrt{K_ZP\widehat h}$. In dimensionful wall formulas, replace $\sqrt2/12$ by $P$ and $(a,b,h)$ by $((\lambda_4+\lambda_5)/\lambda,(\lambda_4-\lambda_5)/\lambda,\widehat h)$; $\int\!\dd z\,(q_g')^2=15.4749\,\mathrm{GeV}$.

\Needspace{6\baselineskip}\section{Monopole stability and magnetic flux}\label{app:orientation}\label{app:flux}
We derive the theory on a fractional string, establish kink stability, and match its magnetic charge to the deconfined wall vortex.

\noindent\textit{Theory on a fractional string.} The coordinate $x$ crosses the string within the interface, while $y$ runs along it. The second effective theory follows by integrating the fractional-string profile across $x$. For Eq.~(\ref{eq:S2}), $\int\dd x\sin^2\theta=2/m_B$ and $\int\dd x\sin\theta=\pi/m_B$. The kinetic density $K_g\sin^2\theta(\partial_\beta\boldsymbol n)^2$ gives $G=2K_g/m_B$. The potential density $\sqrt2[an_3^2\sin^2\theta-2hn_3\sin\theta]/12$ integrates to $\sqrt2 an_3^2/(6m_B)-\pi\sqrt2 hn_3/(6m_B)$, giving Eq.~(\ref{eq:Bpot}). These are leading perturbations about the symmetric host, not exact coefficients at arbitrary $a,h$. The neutral gauge correction is
\begin{equation}
K_Z=K_g-\frac{2g'^2}{g^4}\int\dd z\,(q_g')^2+O(g'^4).
\label{eq:KZexp}
\end{equation}
The profile shift does not enter at this order because $q_g$ already minimizes Eq.~(\ref{eq:K}). Expanding the polar tension $4\sqrt{\sqrt2(-b)K_Z/12}$ gives the positive coefficient $C$ in Eq.~(\ref{eq:Caxis}); the weak gauge source projects onto $n_3$, yielding the anisotropy $-Cn_3^2$.

\noindent\textit{Monopole kink and its stability.} We test the balanced B1 junction using its two disconnected orientation minima. Set $n_3=\cos\alpha$ as in Eq.~(\ref{eq:orient}). For constant $\chi$, subtract the common polar tension and impose $\alpha(-\infty)=0$, $\alpha(+\infty)=\pi$. With $\alpha'=\dd\alpha/\dd y$, its energy $\int\dd y[G(\alpha')^2+D\sin^2\alpha]$ completes to
\begin{equation}
\begin{aligned}
E_M={}&\int\dd y\,(\sqrt G\,\alpha'-\sqrt D\sin\alpha)^2\\
&+2\sqrt{GD}\int_0^\pi\dd\alpha\sin\alpha\ge4\sqrt{GD}.
\end{aligned}
\end{equation}
We next test local stability using the kink Hessian. Equation~(\ref{eq:monokink}) saturates the bound. Its polar Hessian is
\begin{equation}
\mathcal O_M=A_M^\dagger A_M,\quad
A_M=\partial_y+m_M\tanh[m_M(y-y_0)].
\end{equation}
Here $m_M\equiv\sqrt{D/G}$ is the inverse monopole width along the string, in the $y$ direction within the interface. It has a translation zero mode and no negative eigenvalue. Azimuthal variations have nonnegative energy $G\sin^2\alpha(\partial_y\chi)^2$; at finite $e$ their gauge identification and the photon must be retained. Thus the monopole is a topological kink of the string-orientation theory, on a topological wall kink. This second reduction requires $m_M\ll m_B\ll M_\perp$. D1 uses its exchanged coefficients. In B2 unequal pole energies prevent a static infinite-string junction; its acceleration is a tension imbalance, not decay of the balanced kink.

\noindent\textit{Flux and electromagnetic charge.} We match the fluxes in the confined and deconfined configurations. The kink's phase change gives $\Phi_Z=-2\Delta\varphi/g_Z$. At a B1/D1 junction the two outward half-unit branches sum to $4\pi/g_Z$; the unequal B2 fractions have the same sum. In E, $\varphi=\vartheta\equiv\arg(x+iy)$ outside the charged core, and $Z_\vartheta\dd\vartheta=-2q_Z(z)\dd\vartheta/g_Z$. At radius $\rho=\sqrt{x^2+y^2}$,
\begin{equation}
B_\rho^Z=\frac{2q_Z'(z)}{g_Z\rho},\qquad
\int\dd z\,\dd\vartheta\,\rho B_\rho^Z=\frac{4\pi}{g_Z}.
\end{equation}
The same flux is therefore spread radially within the wall. On an enclosing surface avoiding the charged core, regular hypercharge implies $0=\int F^Y=c_W\Phi_{\EM}-s_W\Phi_Z^{\rm out}$, giving Eq.~(\ref{eq:charge}) in both regimes. This fixes the total magnetic charge, not the near-core angular distribution. A full coupled three-dimensional monopole Hessian and field profile are beyond these successive effective-theory results.

\section{Wall-vacuum classification}\label{app:vacua}
We derive the minimization underlying the wall-vacuum table in Fig.~1(a).
We use the reference units $m^2=\lambda=1$ for the wall theory and the field and parameter conventions of the main text:
\begin{equation}
U=\begin{pmatrix}u_0-iu_3&c\\-c^*&u_0+iu_3\end{pmatrix},\qquad u_0^2+u_3^2+|c|^2=1.
\label{sm:eq:U}
\end{equation}
\begin{equation}
\begin{gathered}
V_w=\frac{\sqrt2}{12}(au_0^2+bu_3^2-2hu_0),\\
a=\lambda_4+\lambda_5,\quad b=\lambda_4-\lambda_5,\quad h=6m_{12}^2.
\end{gathered}\label{sm:eq:SM-potential}
\end{equation}

For $h\ge0$, minimize Eq.~(\ref{sm:eq:SM-potential}) at fixed $u_0$. The remaining length $1-u_0^2$ lies in $u_3$ for $b<0$, in $c$ for $b>0$, or in either direction for $b=0$. Thus the minimization reduces to
\begin{equation}
\min_{|u_0|\le1}\big[(a-\min(b,0))u_0^2-2hu_0+\min(b,0)\big].
\label{sm:eq:SM-vacuum-minimization}
\end{equation}

The A and D vacua are $(u_0,u_3,c)=(1,0,0)$ and $(\pm1,0,0)$, respectively. B has $c=0$, $u_0=h/(a-b)$, and $u_3=\pm\sqrt{1-u_0^2}$. C has $u_0=h/a$, $u_3=0$; F has $u_0=h/a$; E has $c=0$; G has $u_3=0$; the remaining components obey Eq.~(\ref{sm:eq:U}). H leaves all of $S^3$. These are the sets before electromagnetic identification: C is a charged circle, E a neutral circle. At $a-b=h>0$ or $a=h>0$, the corresponding B or C/F set collapses to the A point with a vanishing quadratic curvature. A positive common rescaling of $(a,b,h)$ leaves the minimizers unchanged, giving the normalized disk in Fig.~1 of the main text; H is its separate, unnormalized parameter origin. The mixing $h$ is distinct from the bulk bias $\Delta V$.

\section{Coexisting inert and inert-like bulk vacua}\label{sm:app:bias}
This section gives the tree-level bulk-vacuum conditions used in the main text, followed by their continuation to small mixing and the qualifications needed at finite temperature.

\subsection{Extrema, depths, and all scalar curvatures}
Use the potential and normalization of the main text,
\begin{equation}
\begin{aligned}
V={}&-m_1^2r_1-m_2^2r_2+\frac{\lambda_1}{2}r_1^2
+\frac{\lambda_2}{2}r_2^2+\lambda_3r_1r_2\\
&+\lambda_4|\zeta|^2+\lambda_5\operatorname{Re}\zeta^2,
\end{aligned}
\label{sm:eq:sm-display-104}
\end{equation}
where $r_i\equiv H_i^\dagger H_i$, $\zeta\equiv H_1^\dagger H_2$, and all parameters are real. Unlike conventions with $H_i=(0,v_i/\sqrt2)^T$, our $v_i$ denotes the doublet amplitude itself. With $m_i^2>0$ and $\lambda_i>0$, the two axis extrema are
\begin{equation}
\begin{aligned}
\mathrm I:&\quad (H_1,H_2)=(v_1\xi_0,0),\quad v_1^2=m_1^2/\lambda_1,\\
\mathrm{IL}:&\quad (H_1,H_2)=(0,v_2\xi_0),\quad v_2^2=m_2^2/\lambda_2,
\end{aligned}
\label{sm:eq:sm-display-105}
\end{equation}
with $\xi_0=(0,1)^T$, and
\begin{equation}
V_{\mathrm I}=-\frac{m_1^4}{2\lambda_1},\quad
V_{\mathrm{IL}}=-\frac{m_2^4}{2\lambda_2},\quad
\Delta V=\frac{m_1^4}{2\lambda_1}-\frac{m_2^4}{2\lambda_2}.
\label{sm:eq:sm-display-106}
\end{equation}
An extremum need not be a minimum. In the inert-like background, write the noncondensed doublet as $H_1=(\chi^+,(\chi_R+i\chi_I)/\sqrt2)^T$. Its canonically normalized fluctuations have
\begin{equation}
\begin{aligned}
M_{\mathrm{IL},+}^2&=-m_1^2+\lambda_3v_2^2,\\
M_{\mathrm{IL},R}^2&=-m_1^2+(\lambda_3+a)v_2^2,\\
M_{\mathrm{IL},I}^2&=-m_1^2+(\lambda_3+b)v_2^2,
\end{aligned}\label{sm:eq:SM-ILmasses}
\end{equation}
where $a\equiv\lambda_4+\lambda_5$, $b\equiv\lambda_4-\lambda_5$. The condensed-doublet radial mass is $2m_2^2$; its three Goldstone directions are gauge directions. The inert-background masses follow by $1\leftrightarrow2$. Thus the charged direction must be checked as well as both neutral directions \cite{ferreira,ginzburg}.

Define
\begin{equation}
\begin{aligned}
\kappa&\equiv\min(\lambda_3,\lambda_3+a,\lambda_3+b),\\
\rho&\equiv\frac{m_1^2/\sqrt{\lambda_1}}{m_2^2/\sqrt{\lambda_2}},\qquad
\mathcal R_\kappa\equiv\frac{\kappa}{\sqrt{\lambda_1\lambda_2}}.
\end{aligned}
\label{sm:eq:sm-display-108}
\end{equation}
The bounded-from-below conditions are $\lambda_{1,2}>0$ and $\kappa>-\sqrt{\lambda_1\lambda_2}$. All inactive-doublet masses at both extrema are positive precisely when
\begin{equation}
\mathcal R_\kappa^{-1}<\rho<\mathcal R_\kappa,
\qquad \mathcal R_\kappa>1.
\label{sm:eq:sm-display-109}
\end{equation}
Minimizing the quartic cross term over relative orientations gives $\kappa r_1r_2$. In this coexistence region its mixed-amplitude Hessian has determinant proportional to $\lambda_1\lambda_2-\kappa^2<0$, so an interior mixed extremum is not a competing minimum. Consequently the desired tree-level region is
\begin{equation}
1<\rho<\mathcal R_\kappa.\label{sm:eq:SM-metastability}
\end{equation}
Here I is the global minimum and IL is a false, locally stable minimum. At $\rho=1$ they are degenerate. At $\rho=\mathcal R_\kappa$ the lightest inactive-doublet mass at IL vanishes; beyond it IL has a negative curvature. This is the boundary of local stability, rather than a criterion for a long tunneling lifetime.

\subsection{A bias deformation of the analytic reference point}
Keep $\lambda_1=\lambda_2=\lambda$ and deform only the quadratic coefficients:
\begin{equation}
m_1^2=m^2(1+\epsilon),\qquad m_2^2=m^2(1-\epsilon).
\label{sm:eq:sm-display-111}
\end{equation}
Then
\begin{equation}
\Delta V=\frac{2\epsilon m^4}{\lambda},\qquad
0<\epsilon<\frac{\kappa-\lambda}{\kappa+\lambda}.
\label{sm:eq:sm-display-112}
\end{equation}
At the reference point $m^2=\lambda=1$, $\lambda_3=3$, $a=b=0$, the false-vacuum range is $0<\epsilon<1/2$. The inactive-doublet masses at I and IL are respectively $2+4\epsilon$ and $2-4\epsilon$; the radial masses are $2(1+\epsilon)$ and $2(1-\epsilon)$. The degenerate reference point therefore lies away from a bulk loss-of-stability boundary. With small nonzero $a,b$, replace $3$ by $3+\min(0,a,b)$ in the bound on $\epsilon$. Using the analytic reference profile and its overlap coefficients perturbatively requires $\epsilon\ll1$, not merely $\epsilon<1/2$.

\subsection{Small mixing and finite-temperature qualifications}
The Josephson mixing $s\equiv m_{12}^2=h/6$ and the bias are independent deformations. At $s\ne0$, neither axis configuration above is an exact stationary point: the term $-s(\zeta+\zeta^*)$ induces a secondary condensate. Away from vanishing curvatures, both extrema continue as nearby minima. For example, in the neutral real branch,
\begin{equation}
\begin{aligned}
\left.H_2\right|_{\mathrm I}&=\frac{s v_1}{M_{\mathrm I,R}^2}\xi_0+O(s^3),\\
\left.H_1\right|_{\mathrm{IL}}&=\frac{s v_2}{M_{\mathrm{IL},R}^2}\xi_0+O(s^3),
\end{aligned}
\label{sm:eq:sm-display-113}
\end{equation}
while the primary amplitudes shift at $O(s^2)$. The leading changes of the depths are
\begin{equation}
\delta V_{\mathrm I}=-\frac{s^2v_1^2}{M_{\mathrm I,R}^2},\qquad
\delta V_{\mathrm{IL}}=-\frac{s^2v_2^2}{M_{\mathrm{IL},R}^2}.
\label{sm:eq:sm-display-114}
\end{equation}
These expansions require small induced-to-primary amplitude ratios and positive curvature margins. Equation~(\ref{sm:eq:SM-metastability}) is the exact tree-level $s=0$ condition; for $s\ne0$ one must compare the continued depths and Hessians. An interchange-preserving mixing does not itself split the depths at the interchange-symmetric point.

Loop corrections shift both depths and the allowed coexistence region \cite{ferreira}. At finite temperature, the relevant tests use the extrema and curvature Hessian of the thermal effective potential. Substituting thermally corrected quadratic parameters in the tree formula is appropriate only in an approximation that retains the same quartic potential form. An intermediate inert-like phase need not survive as a false vacuum at zero temperature. Ginzburg et al. discuss both direct first-order I$_2\to$I$_1$ transitions and routes through a mixed phase where local stability is lost \cite{ginzburg}. Aoki and Shibuya's bubble-nucleation examples concern finite-temperature broken phases, not a requirement of zero-temperature coexistence \cite{aoki}.

\end{document}